\documentclass{article}

\usepackage{arxiv}

\usepackage[utf8]{inputenc} 
\usepackage[T1]{fontenc}    
\usepackage{hyperref}       
\usepackage{url}            
\usepackage{booktabs}       
\usepackage{amsfonts}       
\usepackage{nicefrac}       
\usepackage{microtype}      
\usepackage{siunitx}
\usepackage{graphicx}
\usepackage{natbib}
\usepackage{doi}
\usepackage{placeins}
\usepackage{caption}
\usepackage{graphicx}
\title{On the emergence of criticality for inhalation-driven particle deposition in the anatomical upper airway}

\author{ \href{https://orcid.org/0000-0000-0000-0000}{\includegraphics[scale=0.06]{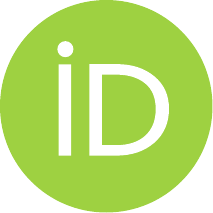}\hspace{1mm}Emma Louwagie}\\
	Department of Mechanical Engineering\\
	South Dakota State University\\
	South Dakota, SD 57006 \\
	\texttt{emma.louwagie@sdstate.edu} \\
	\And
	\href{https://orcid.org/0000-0003-1464-8425}{\includegraphics[scale=0.06]{orcid.pdf}\hspace{1mm}Saikat Basu} \\
	Department of Mechanical Engineering\\
	South Dakota State University\\
	Brookings, SD 57006 \\
	\texttt{saikat.basu@sdstate.edu} \\
}

\renewcommand{\shorttitle}{Self-organized criticality and inhalation-induced upper airway deposition}

\hypersetup{
pdftitle={On the emergence of criticality for inhalation-driven particle deposition in the anatomical upper airway},
pdfsubject={q-bio.NC, q-bio.QM},
pdfauthor={Emma Louwagie, Saikat Basu},
pdfkeywords={First keyword, Second keyword, More},
}

\begin{document}
\maketitle

\begin{abstract}
Inhalation directs air through a defined pathway, initiating from nostrils, moving through the main nasal cavity, past the pharynx and trachea, and culminating in the lungs. Inhaled particles, of a range of sizes, are ferried by this incoming air but are filtered and trapped by upper airway structures to protect the delicate lower respiratory system. From an energetics perspective, the airflow physics along this convoluted tract is characterized by turbulence. The system approaches a critical stationary state over the time scales during which particles enter the airway and deposit. This stasis can be conjectured to correspond with the emergence of criticality in the complex flow domain. For such systemic criticality (i.e., sensitivity to perturbations), inhaled particle deposition impacted by the surrounding flow processes can act as signature `avalanche'-like events. Based on the principles of organized criticality, we have explored the emergence of power law trends in particle deposition levels at the nasopharynx, a key initial infection site for airborne pathogens. These trends are derived from numerical data from five anatomic airway geometries for 15-85 L/min inhalation rates, modeled using high-fidelity Large Eddy Simulations. 
\end{abstract}


\section{Introduction}
Air inhalation through the upper airway follows a direct anatomical pathway within the upper respiratory tract. Upper airway anatomy consists of six specific components: the nasal cavity, the sinus cavities, the pharynx (throat), the larynx (voice box), the bronchi (larger branches of the lungs), and the bronchioles (smaller branches of the lungs). Upon inhalation, air enters the nostrils (or nares), flows into the nasal cavity, and sweeps past the sinuses before reaching the pharynx. Air passes through the first section of the pharynx, the nasopharynx, located posterior to the nose, into the oropharynx, situated directly behind the mouth, and down into the laryngopharynx, located superior to the larynx. From the laryngopharynx, air continues into the larynx, descends through the trachea, and into the bronchi to be distributed throughout the lungs by the bronchioles. The lungs are a very delicate organ that must be protected by many anatomical structures with various functions along the upper airway. Ideal protection mechanisms include filtering and removing harmful debris, pathogens, and particulates. Initial filtering occurs at the cilia or hair in the nares, followed by epithelial cilia and membrane linings within the nasal cavity. Mucosa lines the four paired, air-filled sinuses within the skull bones surrounding the nasal cavity, commonly called the paranasal sinuses, which further contributes to the filtration process. Epithelial membranes lining the nasopharynx filter out a large proportion of the remaining pathogens and particulates. Filtered air is then safe to pass through the remaining structures to complete the inhalation process. Breathing is a critical biological function strongly affected by particles and bacteria \citep{davies2016struct}. Previous research focusing on the breathing process, specifically inhalation, adopts various methodological approaches to develop a deeper understanding of these influences.

\FloatBarrier

\begin{figure}[!htb]
	\centering
        \includegraphics[width=8cm]{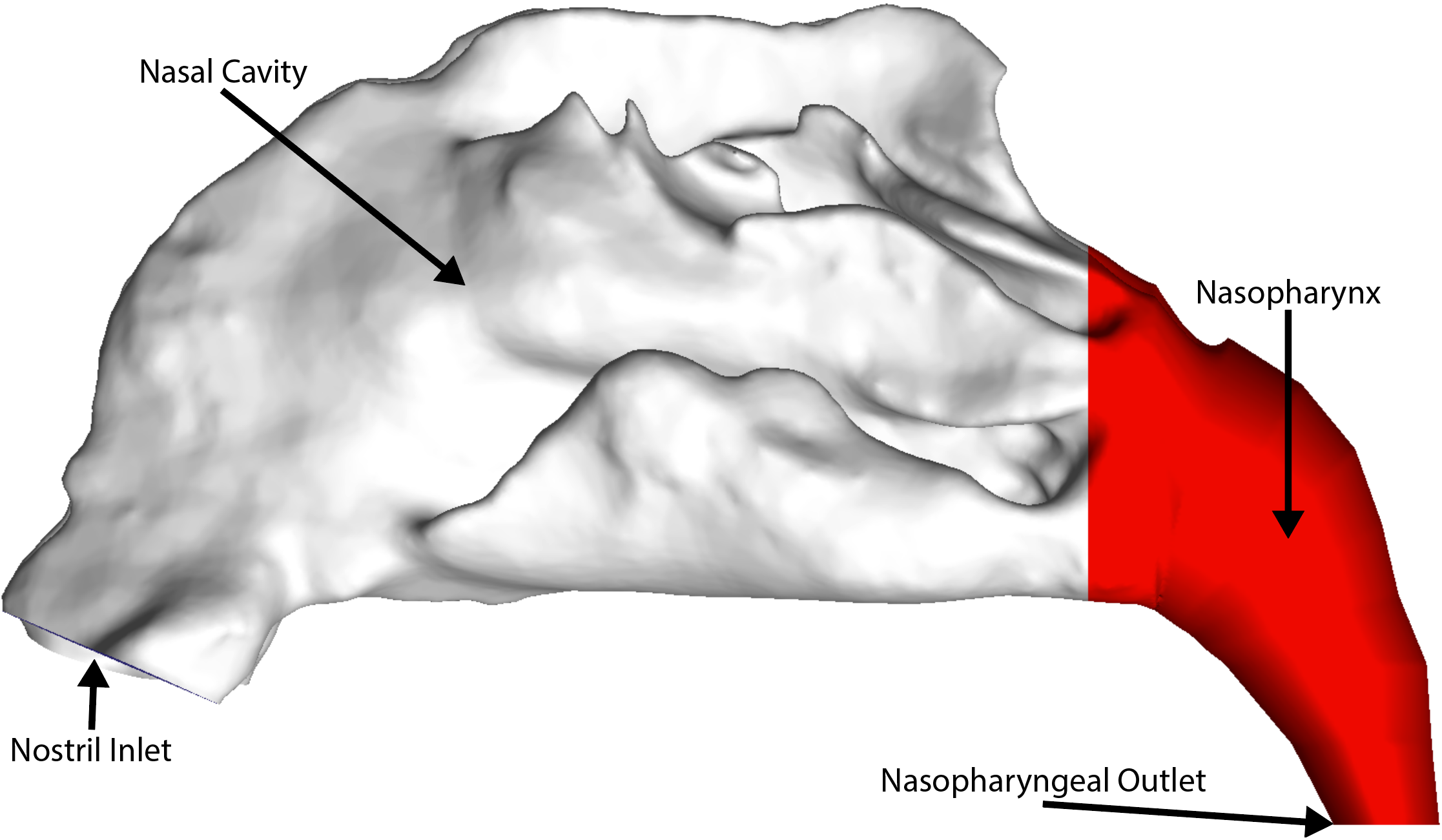}
        \captionsetup{width=7cm}
        \caption{C-T based sample nasal (in grey) and nasopharyngeal (in red) cavities of the first geometry used during the experiment, MU001. Arrows signify important regions utilized during the experiment.}
	\label{fig:fig1}
\end{figure}

\begin{figure}[!htb]
	\centering
	\includegraphics[width=8cm]{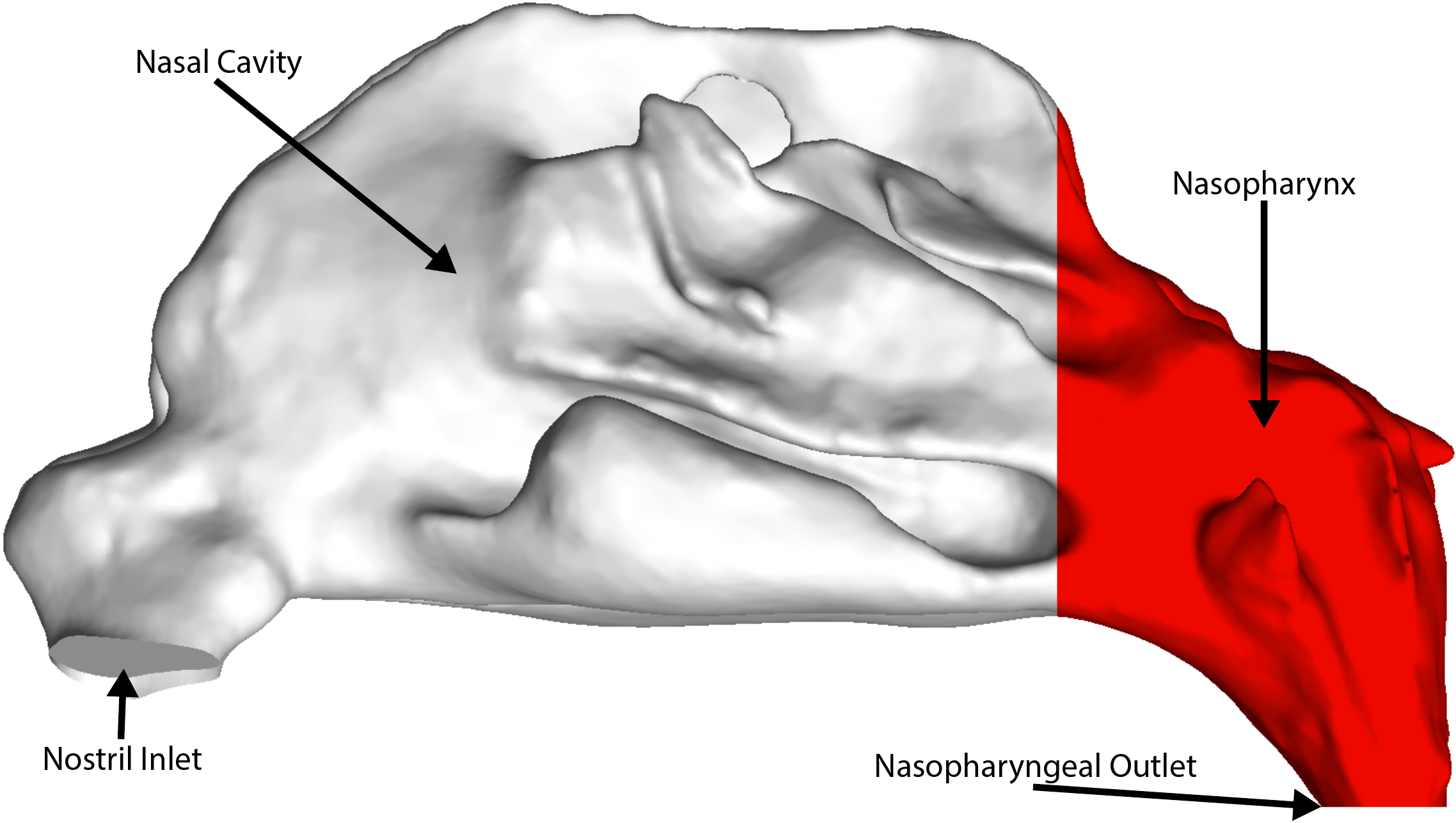}
        \captionsetup{width=7cm}
	\caption{Computed tomography (CT) derived sample nasal (in grey) and nasopharyngeal (in red) cavities of the first geometry used during the experiment, MU004. Arrows signify important regions utilized during the experiment.}
	\label{fig:fig2}
\end{figure}

\begin{figure}[!htb]
	\centering
	\includegraphics[width=8cm]{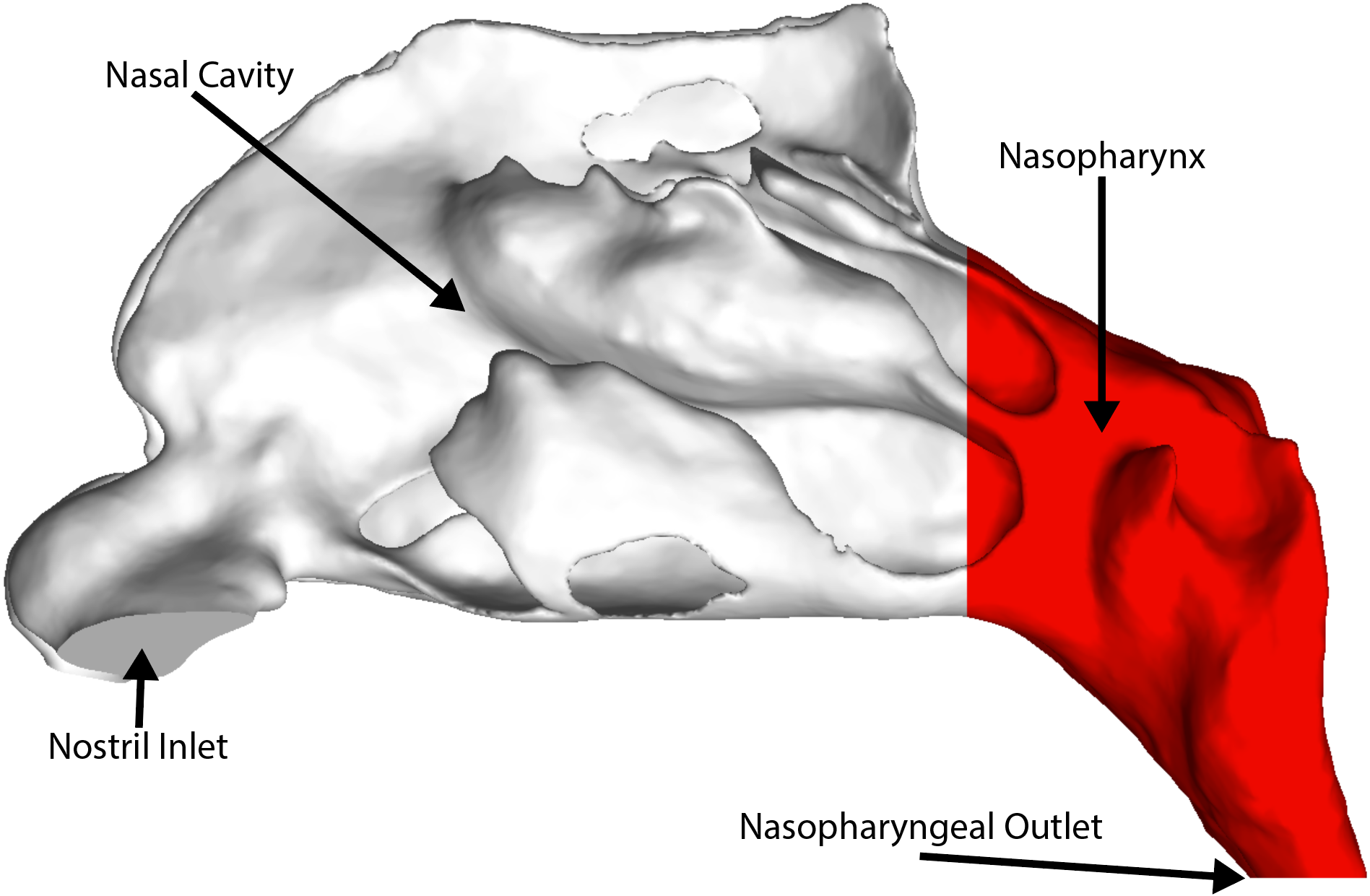}
        \captionsetup{width=7cm}
	\caption{CT-based sample nasal (in grey) and nasopharyngeal (in red) cavities of the first geometry used during the experiment, MU005. Arrows signify important regions utilized during the experiment.}
	\label{fig:fig3}
\end{figure}

\begin{figure}[!htb]
	\centering
	\includegraphics[width=8cm]{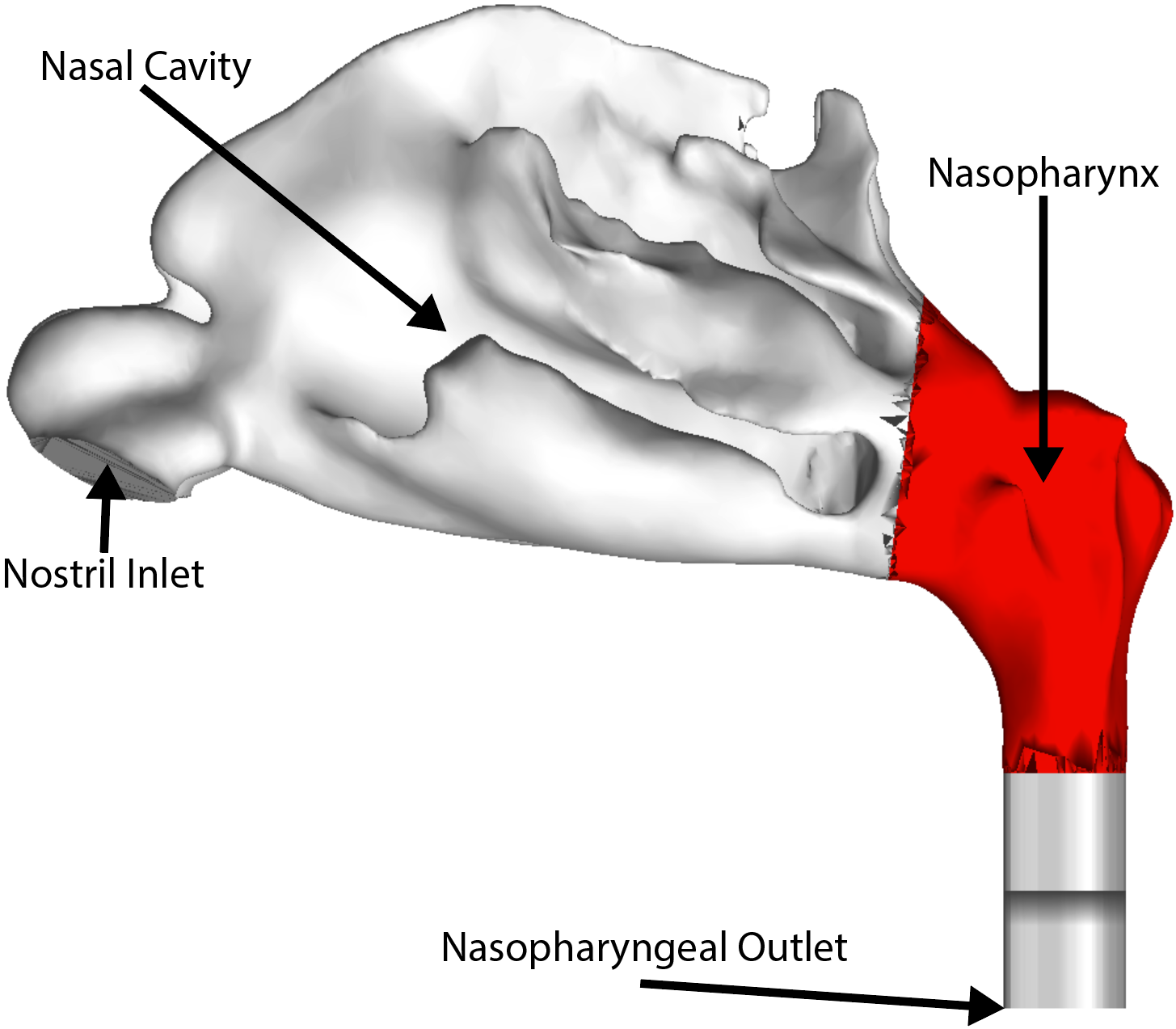}
        \captionsetup{width=7cm}
	\caption{CT-based sample nasal (in grey) and nasopharyngeal (in red) cavities of the first geometry used during the experiment, UNC003. Arrows signify important regions utilized during the experiment.}
	\label{fig:fig4}
\end{figure}

\begin{figure}[!htb]
	\centering
	\includegraphics[width=8cm]{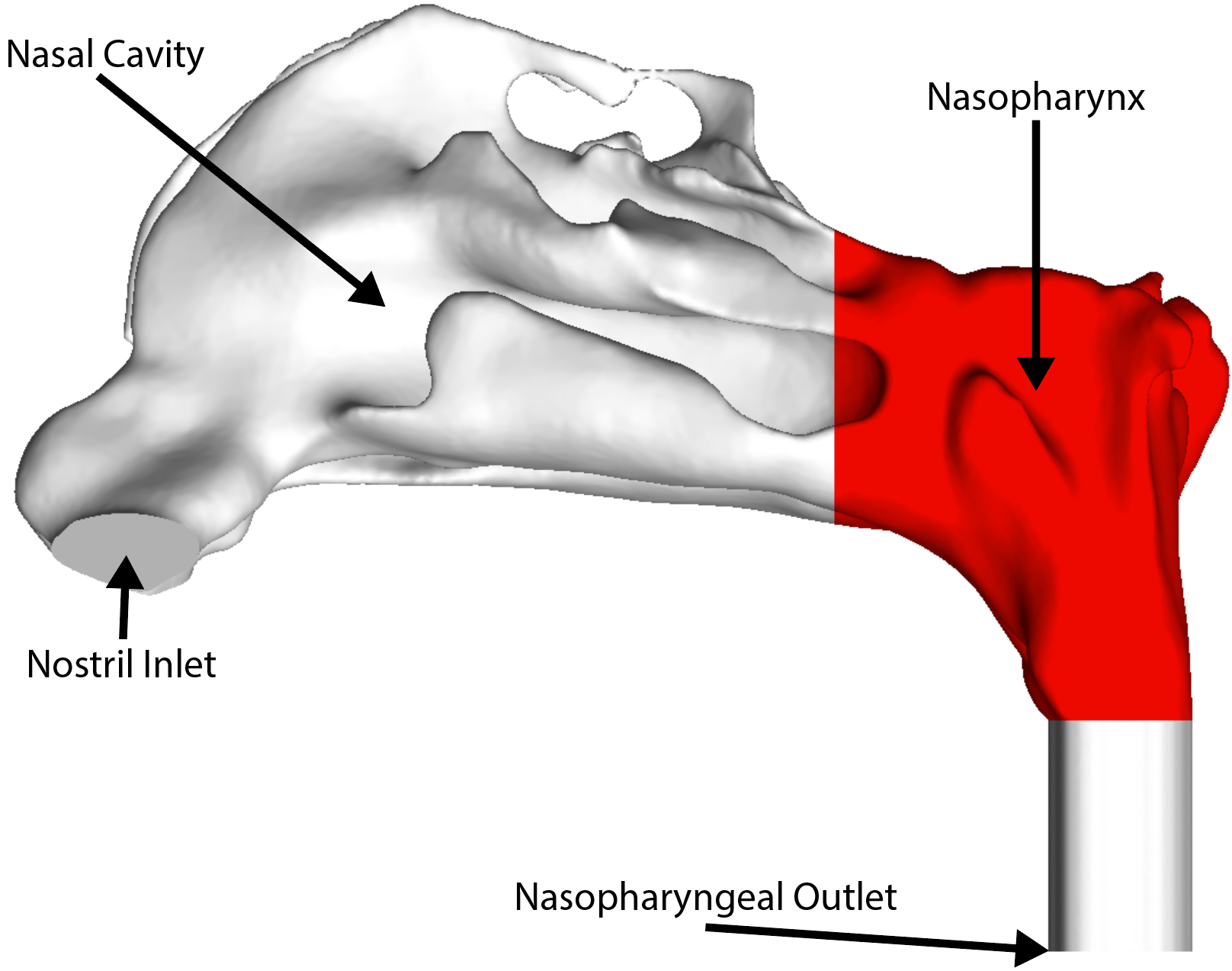}
        \captionsetup{width=7cm}
	\caption{CT-based sample nasal (in grey) and nasopharyngeal (in red) cavities of the first geometry used during the experiment, UNC071. Arrows signify important regions utilized during the experiment.}
	\label{fig:fig5}
\end{figure}

\FloatBarrier

Studies investigating the relationship between inhalation during breathing and mathematical principles utilize experimental data from computational and physical simulations. A study first submitted in 1983 mathematically analyzed tracheobronchial (TB) deposition data by fitting the data into sigmoidal figures to determine the slopes of the quasilinear segments of these curves and the positions of their asymptotes. Results were applied to the evaluation of the health effects of toxic airborne particles \citep{martonen1983dep}. This study showcases an early relationship between a biological function and mathematical principles, inspiring subsequent research. In 1993, a study was published focusing on evaluating potential therapeutic effects of airborne drugs. Researchers recognized the necessity of knowing and understanding deposition sites in the upper airway to determine the effectiveness of aerosolized drug delivery. A mathematical model describing the activity and resulting fate of particles in the lungs was derived using the superposition of distinct processes: inertial impaction, sedimentation, and diffusion of particle deposition. The code necessary to use the mathematical model to calculate totals and compartmental distributions of inhaled aerosol was developed from the experimental data \citep{Martonen1993math}. These two research approaches yielded strong mathematical classifications of breathing dynamics during inhalation. A more recent study conducted in 2018 studied breathing in children as young as five years old to adults. This study specifically focused on existing inhalation therapy for asthma, a globally leading chronic disease. Existing treatment techniques, specifically inhalation therapy, are undergoing studies like this to better understand their effectiveness and potential improvement measures. To simulate inhalation therapy, researchers modeled dry powder inhalations (DPIs) in a realistic model based on the human upper airway \citep{das2018target}. Results indicated that smaller particle sizes correlate with greater deposition at a younger age. Although significant progress has been made on the topic of linking mathematical laws to biological functions, present work has not considered natural organization of deposition patterns within the respiratory tract. Further investigation in this area may develop insights into the predictability of depository behavior.  

Self-organized criticality (SOC) is a mathematical law used to describe complexity through self-organization and criticality. It is considered a naturally occurring phenomenon in which complex systems reach critical points. Various systems exhibit self-organized criticality, particularly those in which small changes lead to significant events \citep{bak1996nature}. This law demonstrates that systems with differing underlying mechanisms exhibit similar critical behavior and statistical properties. Scaled behavior and power-law distributions are used to describe these systems, facilitating a deeper understanding of such phenomena at a fundamental level and allowing for better behavior modeling and analysis \citep{bak1987self}. The power law describing self-organized systems can be described as \citep{bak1991SOC}:
\begin{equation}
	N(s) \:  = \: s^{-\tau}.
\end{equation} 

Here $N(s)$ represents the frequency of events of size $s$ and $\tau$ is the slope of the straight line on a log-log plot of event frequency versus event size. This equation highlights the relationship between event size and occurrence probability in SOC systems.

\section{Materials and Methods}
\label{sec:matandmet}


\subsection{Basic Sandpile Model}
SOC was originally demonstrated by the sandpile model, a simple representation of complex system behavior. The behavior of a sandpile was experimentally studied using a square grid of boxes in which particles were randomly dropped. As the particles in one box reach critical height, the stack topples, avalanching the particles to the neighboring boxes. Avalanche size correlates to a power-law frequency-area distribution with a slope of approximately unity. The slope describes the self-organizing behavior of complex systems to the natural critical state \citep{bak1996nature}. The theory of SOC proposes the idea of sandpiles occurring as naturally occurring self-organizing phenomena.

Similarly, this study proposes particle deposition in the nasopharynx during inhalation exhibits analogous self-organizing behavior. To support this idea, a comparison between self-organization in sandpiles and nasopharyngeal deposition could be invoked. For example, a sandpile is a buildup of individual grains of sand that begin to topple or avalanche down the edges of the sandpile as they approach critical height. Similarly, in the nasopharynx, epithelial deposition sites trap and collect particles until the site reaches capacity and begins to topple, or avalanche, towards neighboring sites. Computational simulations were conducted to validate the theory of self-organization during inhalation. These simulations modeled particle deposition behavior in the nasopharynx, demonstrating parallel behavior as displayed in the sandpile model.

\subsection{Numerical Simulations}
To investigate particle deposition behavior in the nasopharynx during inhalation, computational fluid dynamics (CFD) simulations were performed; see \citep{basu2018ijnmbe,basu2020scirep,basu2021scirep,akash2023fdd} for prior publications from us on the methodological details. Anatomical segmentation and reconstructions of the CT-derived human airways was done on Mimics and utilized for flow simulation in subjects to ensure anatomical accuracy. The ANSYS software package integrating computer engineering and manufacturing (ICEM CFD) was applied to conduct meshing processes. Each geometry underwent tetrahedral meshing followed by prism meshing along the cavity walls, resulting in meshes with approximately 3 to 5 million tri mesh elements, 1 to 2 million penta mesh elements, 1 to 2 thousand quad mesh elements, and 1 to 2 thousand pyra mesh elements. Completed meshes were imported into ANSYS Fluent for CFD simulations across four flow rates: 15, 30, 55, and 85 L/min. Large Eddy Simulation (LES) with the Kinetic Energy Transport Method (KETM) was the primary flow model and sub-grid scale model utilized during simulations. Time-related parameters include a time set of 0.0002 seconds with total time steps equal to 1,750 \citep{xu2020invest}. ANSYS Fluent’s discrete phase model (DPM) software was used for particle tracking using particles sized from 1 micron to 30 microns, incrementing by 1 micron each time. Deposition patterns were analyzed using the discrete phase model with a Runge-Kutta high-order and implicit low-order scheme. Air density was set to 1.204 kg/m\textsuperscript{3} to replicate the warmed-up air breathed into the nasopharynx. Nostrils were assigned as injection sites during particle tracking simulations to replicate their behavior during inhalation \citep{das2018target}. Data collected during CFD simulations was graphically analyzed to provide insights into deposition behavior within the nasopharynx.

\subsection{Data Transformation}
Data transformation is an important aspect of self-organized criticality evaluation as it employs a power law. Logarithmic scaling is necessary to visualize power law distributions through straight-line transformation. Representing the data as a straight line facilitates pattern identification and analysis. In this study, logarithmic base 10 scaling was applied to particle deposition percentage in the nasopharynx. This scaling technique allows for the compression of wide dataset ranges into a more manageable scale that can include small and large events on the same graph, simplifying the process and improving the accuracy of slope determination. SOC analysis utilizes critical exponents to describe behavior near critical points or heights, such as critical particle deposition thresholds \citep{bak1991SOC}. Logarithmic scaling was used to create a linear replication of the relationship between data points and variables to extract these exponents. Large datasets commonly contain noise, which can be significantly reduced by logarithmic scaling, enabling a clearer understanding of underlying trends and behaviors. In this experiment, applying logarithmic base 10 scaling to the data produced graphs demonstrating linear trends, confirming the presence of SOC.

\begin{figure}[!htb]
	\centering
	\includegraphics[width=12cm]{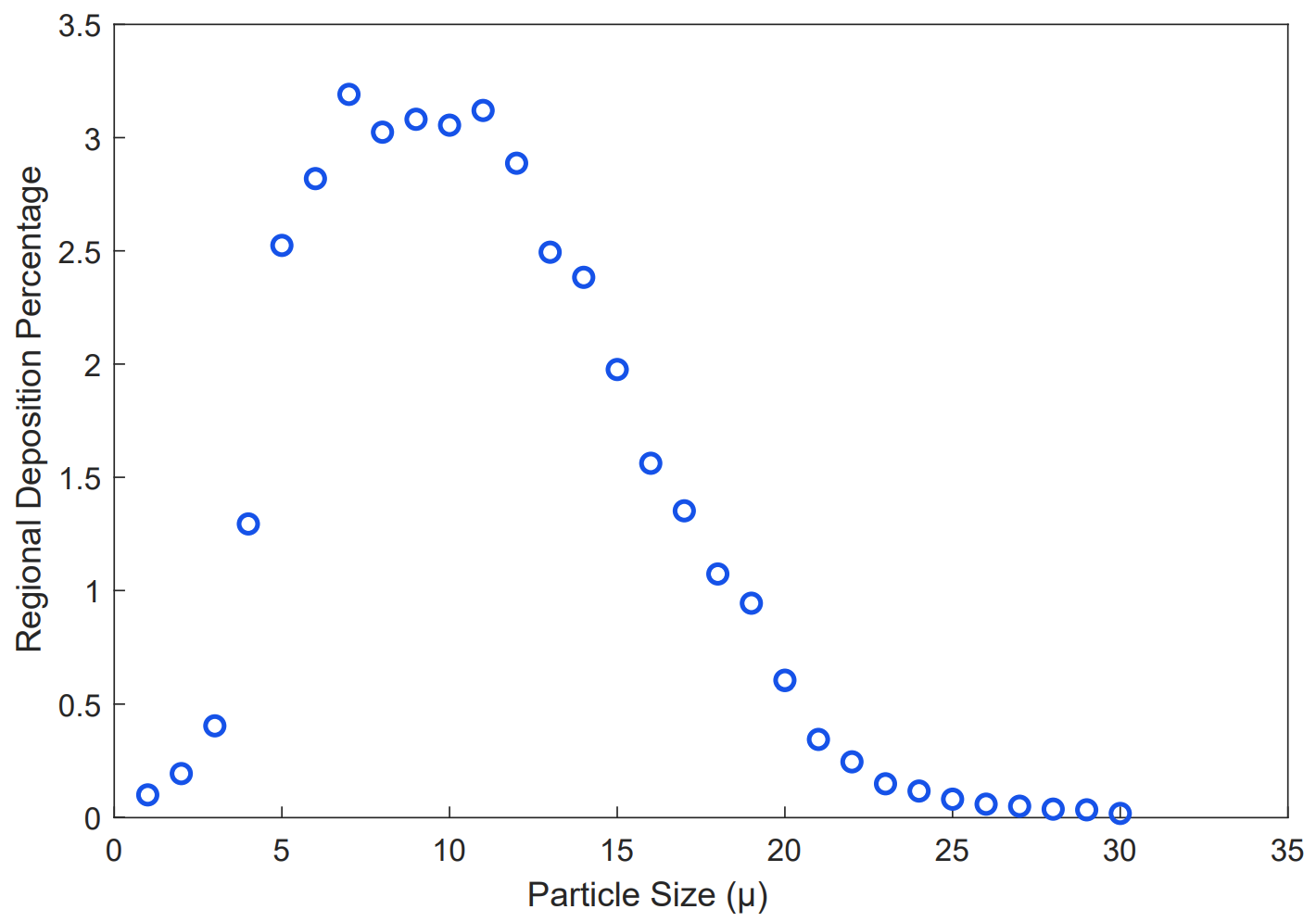}
        \captionsetup{width=10cm}
	\caption{\footnotesize Particle tracking at sizes from 1 micron to 30 microns during inhalation in the nasopharynx resulted in averages that graphically formed a bell shape when regional deposition percentage was graphed against particle size. Averages were taken across all five geometries at all four flow rates: 15, 30, 55, and 85 L/min. According to the graph, the critical point occurs at approximately 7 microns. }
	\label{fig:fig6}
\end{figure}

\begin{figure}[!htb]
	\centering
	\includegraphics[width=12cm]{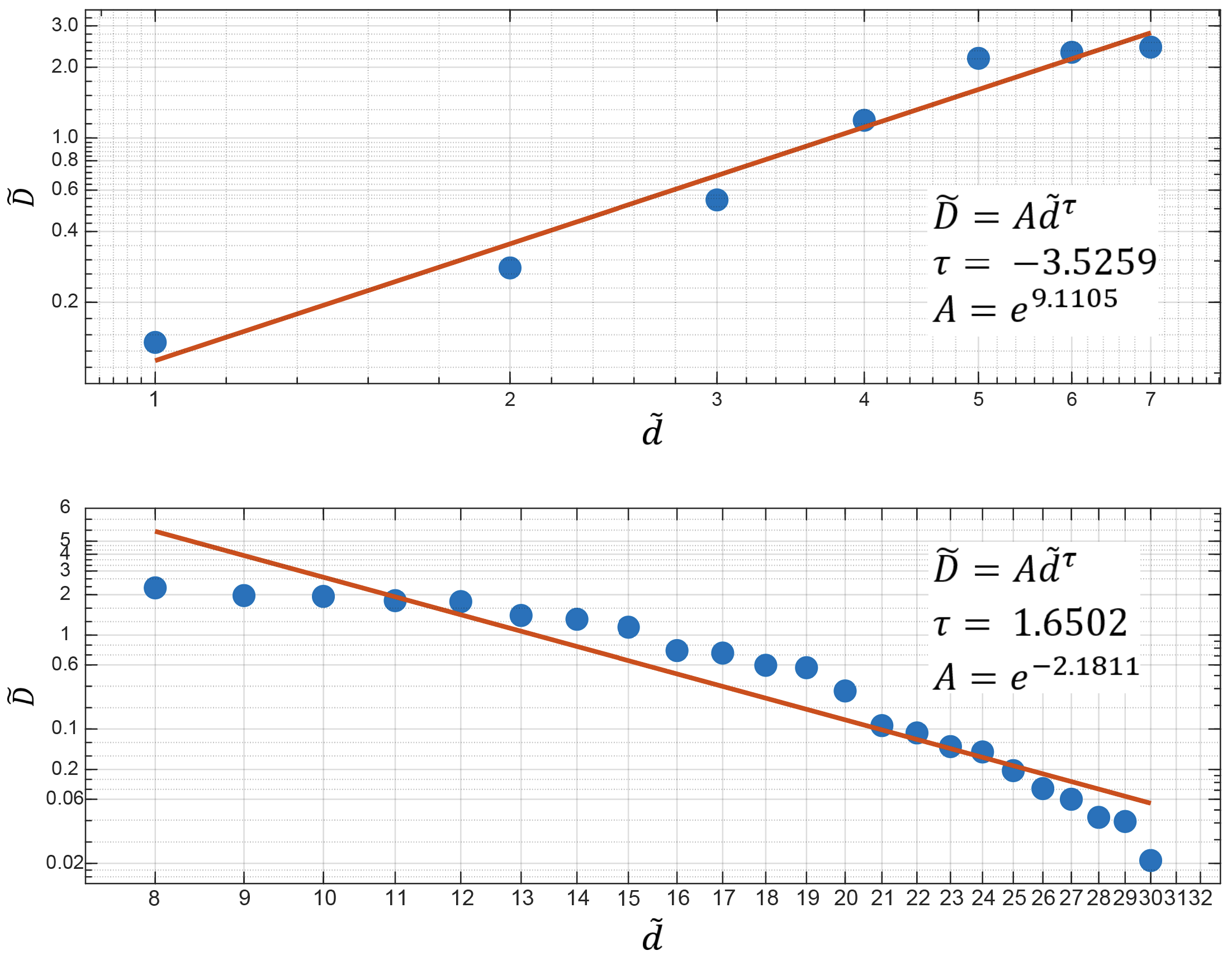}
        \captionsetup{width=10cm}
	\caption{Resulting power law trend extracted from flow-averaged nasopharyngeal deposition data in five anatomic geometries. The inhaled airflow was simulated at 15, 30, 55, and 85 L/min. Deposition efficiency at the nasopharynx increased linearly on a log-log scale for particle sizes 1 -- \SI{7}{\micro\metre} and then declined linearly on a log-log scale for particle sizes 8 -- \SI{30}{\micro\metre}. Here $\tilde{\mathcal{D}} = \log_e\left[\,\langle \mathcal{D} \rangle \,\right]$, with $\mathcal{D}$ being the percentage of tracked particles undergoing nasopharyngeal deposition. $\tilde{\mathcal{\delta}} = \log_e\left[\,\delta\,\right]$, with $\delta$ being the tracked particle sizes. The values of the critical exponents $\tau$ fit with the SOC-prescribed typical range.}
	\label{fig:fig7}
\end{figure}

\FloatBarrier

\section{Discussion}
\label{sec:discussion}
Self-organized criticality requires the system to exhibit behavior independent of the observed scale, meaning the occurrence of small-scale events is equally as likely as the occurrence of large-scale events within the system. The system must naturally reach a critical state, where the power law can be observed, without any external interventions. Time scales representing the driving force, and the internal dynamics of the system must exist on separate time scales, as the driving force will be slower than the rapid internal dynamics. Constant driving input will prevent the system from existing in a state of equilibrium. Power laws are considered universal, and do not depend on specific and intricate details of the system but instead on the general dynamics of the system. In this study, small-scale events during particle deposition in the nasopharynx were observed to be equally as common as large-scale events. Graphical trends were similar across all 30 particles sizes at all four flow rates, in which deposition reached a critical state within 3 to 21 microns for each anatomical geometry. Before reaching critical height, deposition followed an increasing trend, and after surpassing critical height, the deposition followed a decreasing trend. The driving force in this scenario is represented by the inhalation flow rate existing on a slower time scale than the avalanches occurring within the nasopharynx, which occurs within the time-step of 0.0002 seconds. Across 20 experiments, utilizing five anatomical geometries and four flow rates, consistent critical heights and power-law trends were observed, demonstrating the universality of SOC.  Self-organized criticality requires a slope with a critical exponent similar to or between the values of 1 and 3. In these experiments, particle deposition demonstrated an increasing linear trend with a power-law slope of 1.6502 from 1 micron to 7 microns. After reaching the critical point of 7 microns, deposition followed a decreasing linear trend until 30 microns, with a power law slope of -3.529. Graphs showcasing the averaged data with logarithmic base 10 scaling from these experiments can be found in Figures 1 and 2. The constant exponential values were calculated using the power law governing self-organized criticality, and fall near or within the required range, further supporting the presence of self-organized criticality in nasopharyngeal particle deposition.

\begin{figure}[!htb]
	\centering
	\includegraphics[width=11cm]{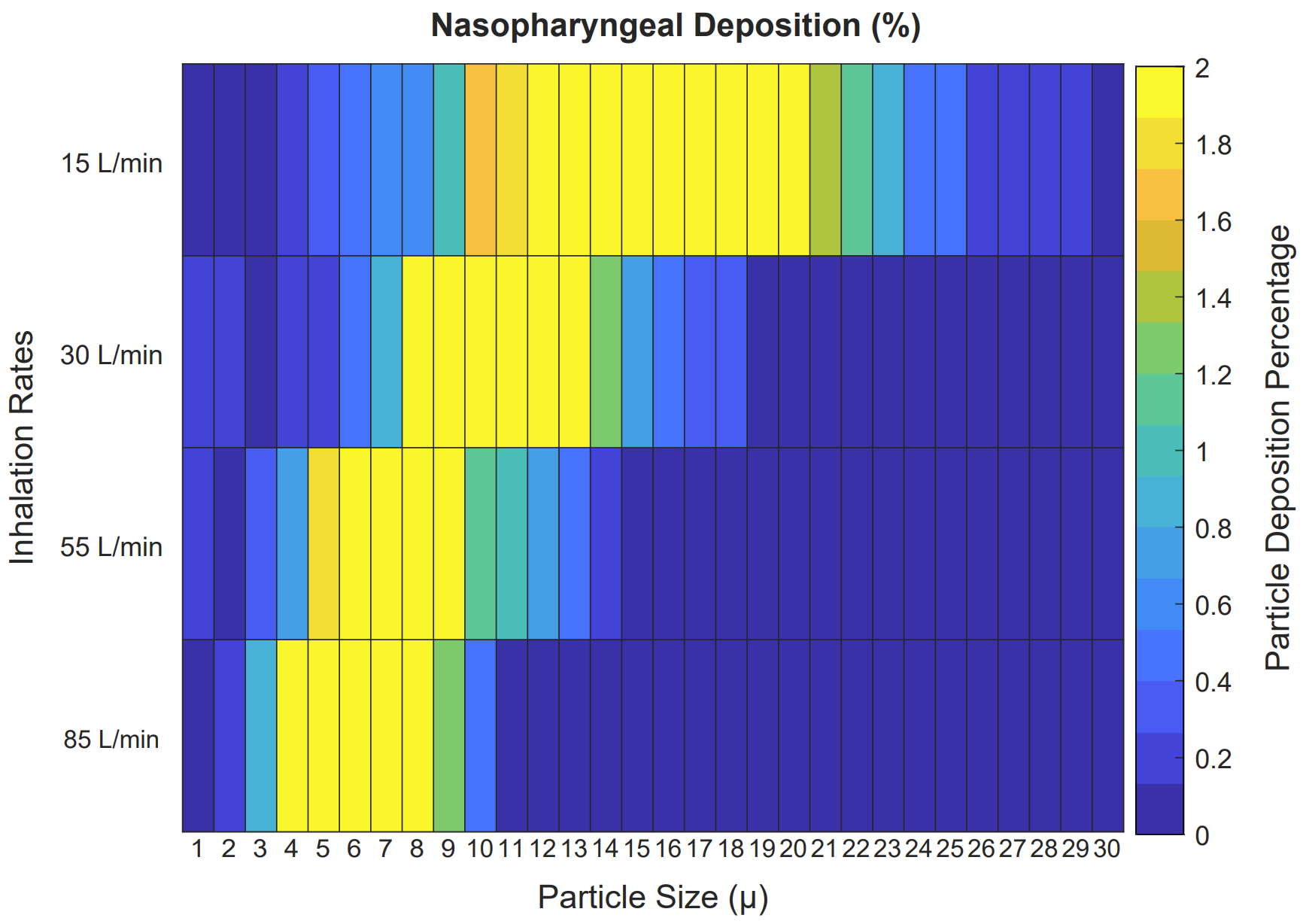}
        \captionsetup{width=10cm}
	\caption{Nasopharyngeal particle deposition in MU001 is presented in color map format. The horizontal rectangles represent the four flow rates, and the vertical rectangles represent particle sizes. Particle deposition percentage is represented by the color scale on the right side, with dark blue indicating a deposition percentage of 0 and bright yellow representing a deposition percentage of 2. The highest percentages are centered between 3 and \SI{20}{\micro\metre}, indicating a critical point within that range.}
	\label{fig:fig8}
\end{figure}

\begin{figure}[!htb]
	\centering
	\includegraphics[width=11cm]{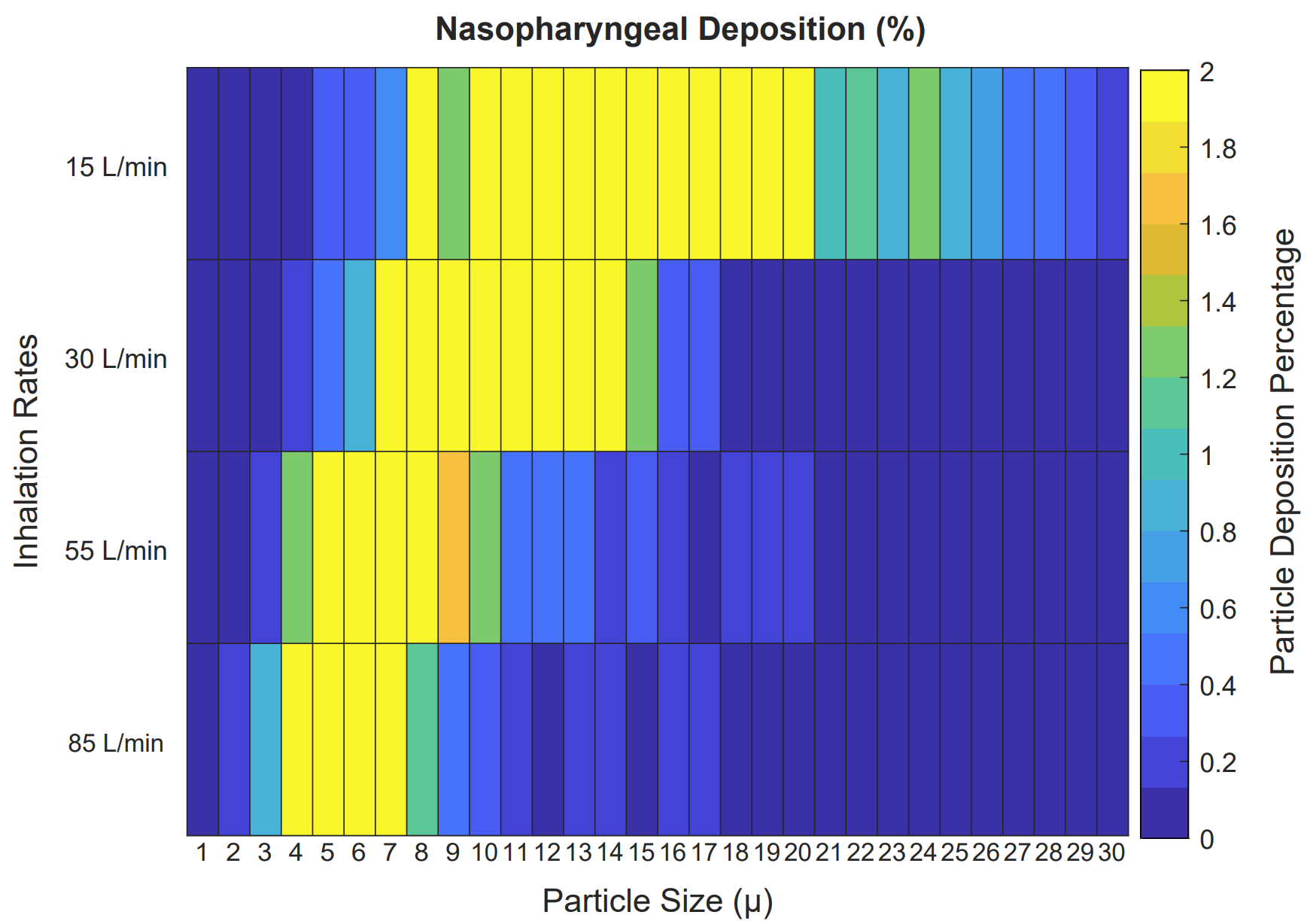}
        \captionsetup{width=10cm}
	\caption{Nasopharyngeal particle deposition in MU004 is presented in color map format. The horizontal rectangles represent the four flow rates, and the vertical rectangles represent particle sizes. Particle deposition percentage is represented by the color scale on the right side, with dark blue indicating a deposition percentage of 0 and bright yellow representing a deposition percentage of 2. The highest percentages are centered between 4 and \SI{20}{\micro\metre}, indicating a critical point within that range.}
	\label{fig:fig9}
\end{figure}

\begin{figure}[!htb]
	\centering
	\includegraphics[width=11cm]{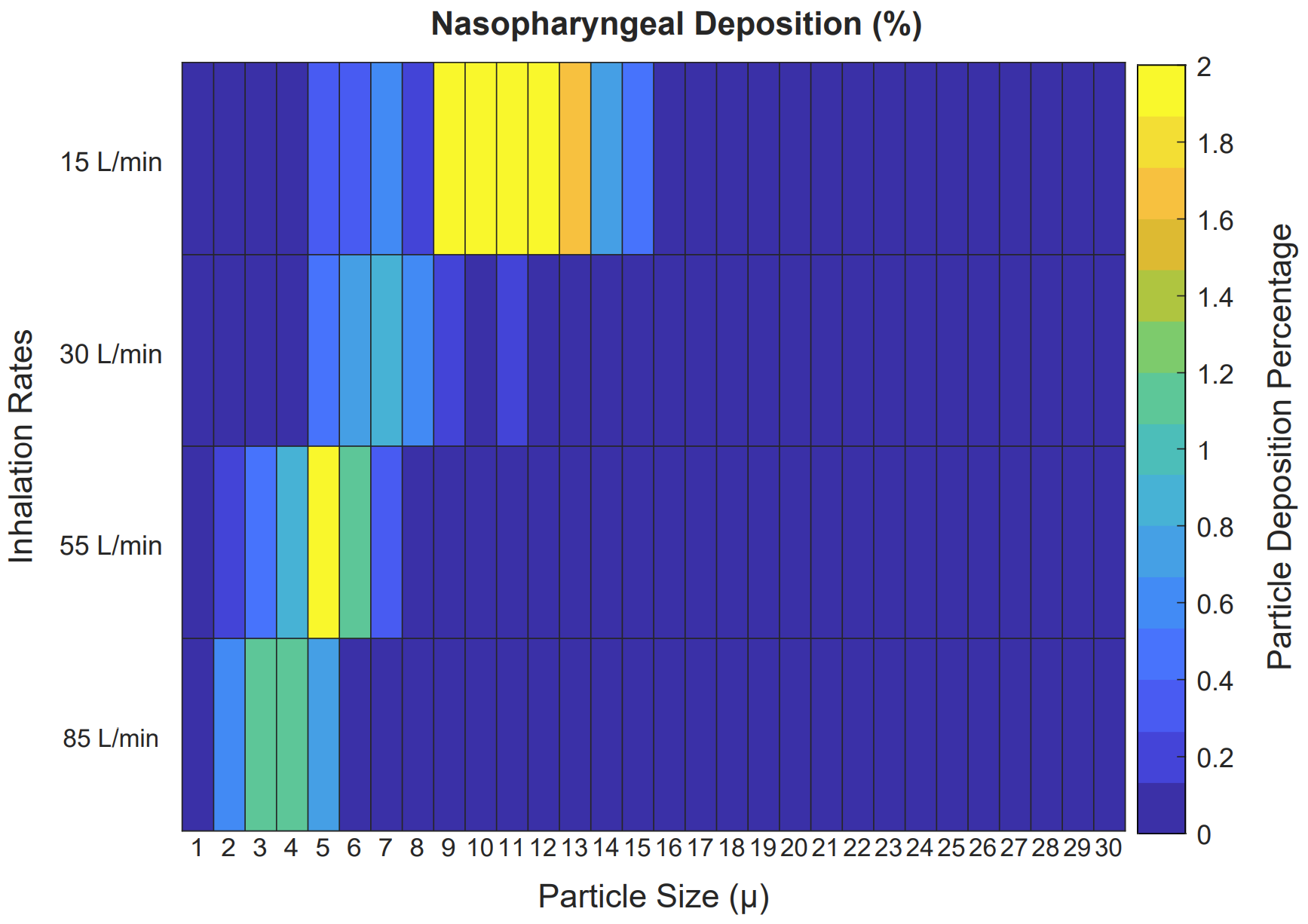}
        \captionsetup{width=10cm}
	\caption{Nasopharyngeal particle deposition in MU005 is presented in color map format. The horizontal rectangles represent the four flow rates, and the vertical rectangles represent particle sizes. Particle deposition percentage is represented by the color scale on the right side, with dark blue indicating a deposition percentage of 0 and bright yellow representing a deposition percentage of 2. The highest percentages are centered between 5 and \SI{12}{\micro\metre}, indicating a critical point within that range.}
	\label{fig:fig10}
\end{figure}

\begin{figure}[!htb]
	\centering
	\includegraphics[width=11cm]{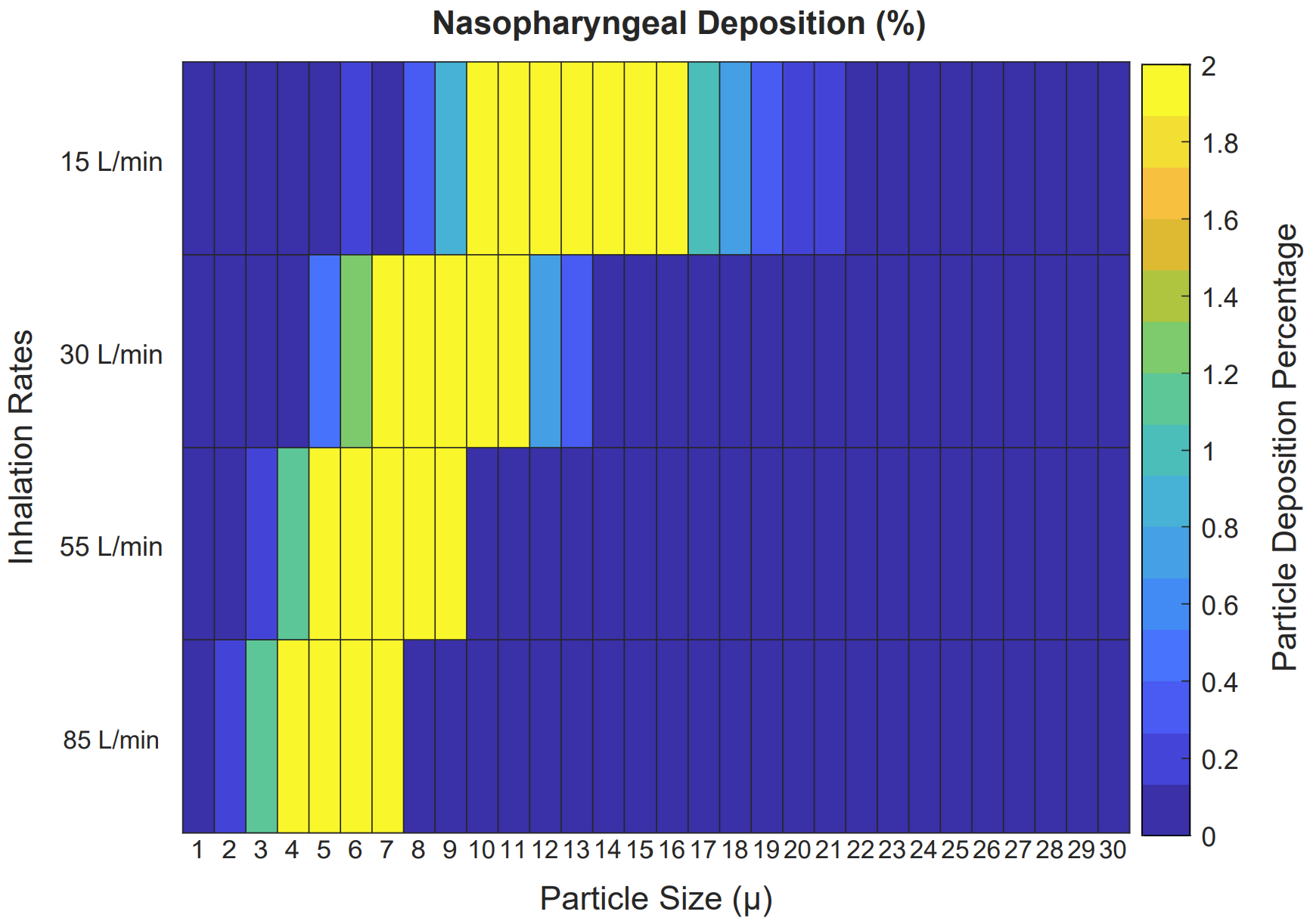}
        \captionsetup{width=10cm}
	\caption{Nasopharyngeal particle deposition in UNC003 is presented in color map format. The horizontal rectangles represent the four flow rates, and the vertical rectangles represent particle sizes. Particle deposition percentage is represented by the color scale on the right side, with dark blue indicating a deposition percentage of 0 and bright yellow representing a deposition percentage of 2. The highest percentages are centered between 4 and \SI{16}{\micro\metre}, indicating a critical point within that range.}
	\label{fig:fig11}
\end{figure}

\begin{figure}[!htb]
	\centering
	\includegraphics[width=11cm]{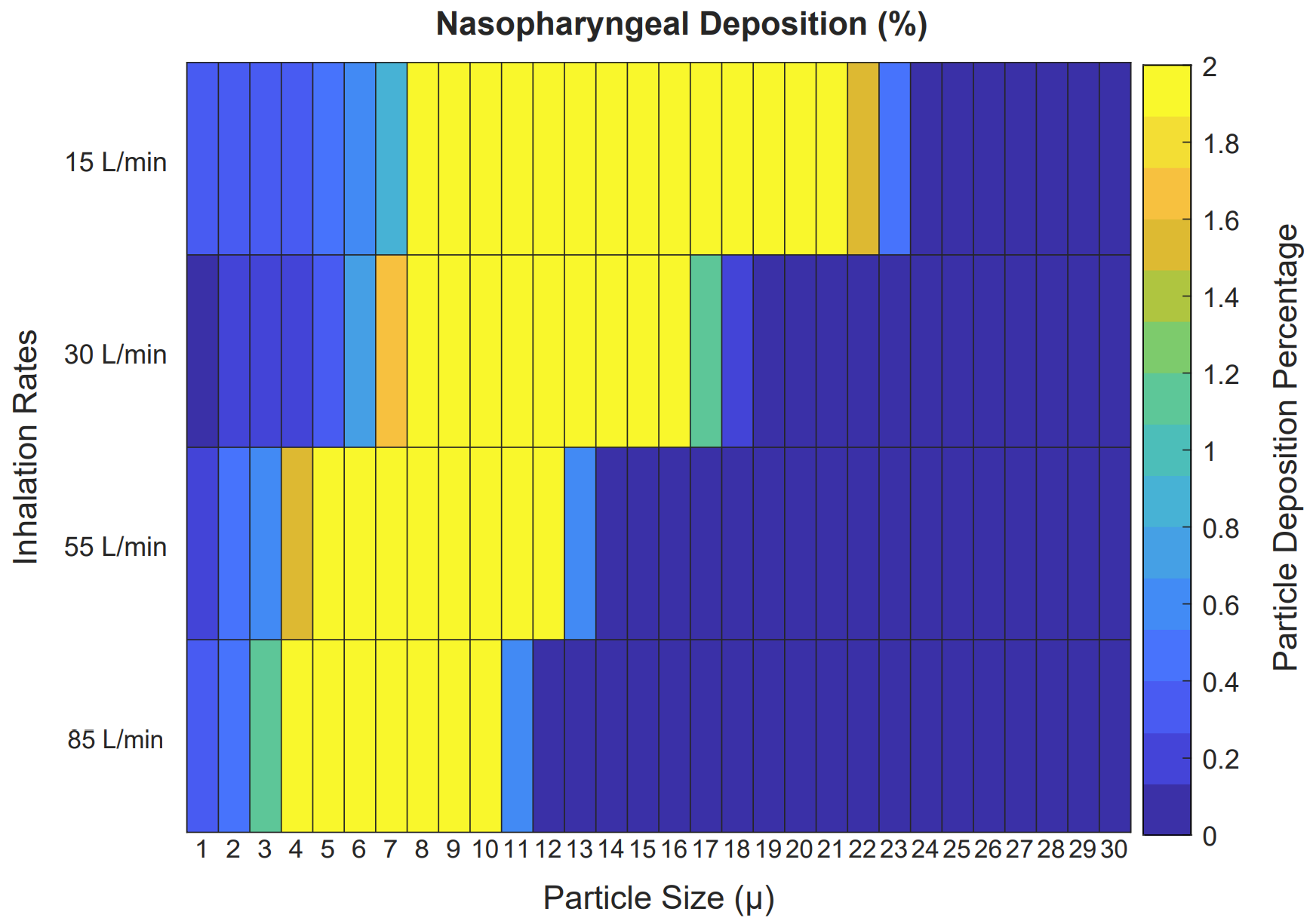}
        \captionsetup{width=10cm}
	\caption{Nasopharyngeal particle deposition in UNC071 is presented in color map format. The horizontal rectangles represent the four flow rates, and the vertical rectangles represent particle sizes. Particle deposition percentage is represented by the color scale on the right side, with dark blue indicating a deposition percentage of 0 and bright yellow representing a deposition percentage of 2. The highest percentages are centered between 4 and \SI{21}{\micro\metre}, indicating a critical point within that range.}
	\label{fig:fig12}
\end{figure}


\bibliographystyle{unsrtnat}
\bibliography{references}  

\begin{thebibliography}{12}
\providecommand{\natexlab}[1]{#1}
\providecommand{\url}[1]{\texttt{#1}}
\expandafter\ifx\csname urlstyle\endcsname\relax
  \providecommand{\doi}[1]{doi: #1}\else
  \providecommand{\doi}{doi: \begingroup \urlstyle{rm}\Url}\fi

\bibitem[Davies(2016)]{davies2016struct}
Moores Davies.
\newblock Structure of the respiratory system, related to function.
\newblock \emph{PubMed Central}, 2016.

\bibitem[Martonen(1983)]{martonen1983dep}
Martonen.
\newblock Deposition of inhaled particulate matter in the upper respiratory tract, larynx, and bronchial airways: A mathematical description.
\newblock \emph{Journal of Toxicology and Environmental Health}, 1983.

\bibitem[Martonen(1993)]{Martonen1993math}
Martonen.
\newblock Mathematical model for the selective deposition of inhaled pharmaceuticals.
\newblock \emph{Journal of Pharmaceutical Sciences}, 1993.

\bibitem[Das et~al.(2018)Das, Nof, Amirav, Kassinos, and Sznitman]{das2018target}
P.~Das, E.~Nof, I.~Amirav, S.~C. Kassinos, and J.~Sznitman.
\newblock Targeting inhaled aerosol delovery to upper airways in children: Insight from computational fluid dynamics (cfd).
\newblock \emph{Plos One}, 2018.

\bibitem[Bak(1996)]{bak1996nature}
Bak.
\newblock \emph{How Nature Work the science of self-organized criticality}.
\newblock Copernicus New York, NY, 1996.

\bibitem[Bak(1987)]{bak1987self}
Wiesenfeld Bak, Tang.
\newblock Self-organized criticality: An explanation of 1/f noise.
\newblock \emph{Physical Review Letters}, 1987.

\bibitem[Bak(1991)]{bak1991SOC}
Chen Bak.
\newblock Self-organized criticality.
\newblock \emph{Scientific American}, 1991.

\bibitem[Basu et~al.(2018)Basu, Frank-Ito, and Kimbell]{basu2018ijnmbe}
S.~Basu, D.~O. Frank-Ito, and J.~S. Kimbell.
\newblock {On computational fluid dynamics models for sinonasal drug transport: Relevance of nozzle subtraction and nasal vestibular dilation}.
\newblock \emph{International Journal for Numerical Methods in Biomedical Engineering}, 34\penalty0 (4):\penalty0 e2946, 2018.

\bibitem[Basu et~al.(2020)Basu, Holbrook, Kudlaty, Fasanmade, Wu, Burke, Langworthy, Farzal, Mamdani, Bennett, Fine, Senior, Zanation, Ebert~Jr, Kimple, Thorp, Frank-Ito, Garcia, and Kimbell]{basu2020scirep}
S.~Basu, L.~T. Holbrook, K.~Kudlaty, O.~Fasanmade, J.~Wu, A.~Burke, B.~W. Langworthy, Z.~Farzal, M.~Mamdani, W.~D. Bennett, J.~P. Fine, B.~A. Senior, A.~M. Zanation, C.~S. Ebert~Jr, A.~J. Kimple, B.~D. Thorp, D.~O. Frank-Ito, G.~J.~M. Garcia, and J.~S. Kimbell.
\newblock {Numerical evaluation of spray position for improved nasal drug delivery}.
\newblock \emph{Scientific Reports}, 10\penalty0 (1):\penalty0 1--18, 2020.

\bibitem[Basu(2021)]{basu2021scirep}
S.~Basu.
\newblock {Computational characterization of inhaled droplet transport to the nasopharynx}.
\newblock \emph{Scientific Reports}, 11:\penalty0 1--13, 2021.

\bibitem[Akash et~al.(2023)Akash, Lao, Balivada, Ato, Ka, Mituniewicz, Silfen, Suman, Chakravarty, Joseph-McCarthy, and Basu]{akash2023fdd}
M.~M.~H. Akash, Y.~Lao, P.~A. Balivada, P.~Ato, N.~K. Ka, A.~Mituniewicz, Z.~Silfen, J.~D. Suman, A.~Chakravarty, D.~Joseph-McCarthy, and S.~Basu.
\newblock {On a model-based approach to improve intranasal spray targeting for respiratory viral infections}.
\newblock \emph{Frontiers in Drug Delivery -- Sec.~Respiratory Drug Delivery}, 3, 2023.
\newblock \doi{10.3389/fddev.2023.1164671}.

\bibitem[Xu et~al.(2020)Xu, Wu, Weng, and Fu]{xu2020invest}
X.~Xu, J.~Wu, W.~Weng, and M.~Fu.
\newblock Investigation of inhalation and exhalation flow pattern in a realistic human upper airway model by piv experiments and cfd simulations.
\newblock \emph{Biomechanics and Modeling in Mechanobiology}, 2020.

\end{thebibliography}






\end{document}